\documentclass[preprint,pra,showpacs,nofootinbib]{revtex4}
\usepackage{mathrsfs}
\usepackage{float,epsfig}
\usepackage{graphicx}
\usepackage{dcolumn}
\usepackage{bm}
\usepackage{amsmath,amssymb,amsthm}
\usepackage[colorlinks=true,linkcolor=blue]{hyperref}
\usepackage{subfigure}
\usepackage{booktabs}
\usepackage[mathscr]{euscript}

\newcommand{\bea}{\begin{eqnarray}}
\newcommand{\eea}{\end{eqnarray}}
\newcommand{\beq}{\begin{equation}}
\newcommand{\eeq}{\end{equation}}
\newcommand{\nn}{\nonumber}
\def\/{\over}
\newcommand{\bra}[1]{\langle#1|}
\newcommand{\ket}[1]{|#1\rangle}

\begin{document}
\title{Entanglement dynamics for uniformly accelerated two-level atoms}
\author{  Jiawei Hu and Hongwei Yu
}
\affiliation{
Center for Nonlinear Science and Department of Physics, Ningbo University,  Ningbo, Zhejiang 315211, China
}

\begin{abstract}

We study, in the paradigm of open quantum systems, the entanglement dynamics of two uniformly accelerated atoms with the same acceleration perpendicular to the separation. The two-atom system is treated as an open system coupled with a bath of fluctuating massless scalar fields in the Minkowski vacuum, and the master equation that governs its evolution is derived. It has been found that, for accelerated atoms with a nonvanishing separation, entanglement sudden death is a general feature when the initial state is entangled, while for those in a separable initial state, entanglement sudden birth only happens for atoms with an appropriate interatomic separation and sufficiently small acceleration. Remarkably, accelerated atoms can get entangled in certain circumstances while the inertial ones in the Minkowski vacuum cannot.  A comparison between the results of accelerated atoms and those of static ones in a thermal bath shows that uniformly accelerated atoms exhibit features distinct from those immersed in a thermal bath at the Unruh temperature in terms of entanglement dynamics.
\end{abstract}

\maketitle

\section{Introduction}

Entanglement is one of the most fascinating features which distinguish the classical and quantum worlds, and it plays the key role in quantum information science \cite{information}. The inevitable interactions between quantum systems and the environment, which cause decoherence and may lead to entanglement degradation, are one of the main obstacles in the realization of quantum technologies. Therefore, the time evolution of quantum entanglement between atoms under the influence of external environment is an important  issue in quantum information science. Recently, it has been found that although local decoherence processes take an infinite time, two atoms may get completely disentangled within a finite time. This phenomenon, named  entanglement sudden death \cite{esd-ty,esd-ty2}, has attracted broad attention \cite{esd-ty,esd-ty2,esd-3,esd-4,esd-5,esd-6,esd-7,esd-8,esd-9,Ali-09}. On the other hand, if the atoms are placed in a common bath, indirect interactions between otherwise independent atoms can be generated through correlations that exist, and this leads to interesting phenomena such as the revival of destroyed entanglement \cite{ent-revival} and the creation of entanglement in initially separable states \cite{ent-cre1,ent-cre2,ent-cre3,Benatti-prl,Benatti-job}. For specific initial states, the entanglement generated by the dissipative evolution may exhibit a delayed feature, which is called the delayed sudden birth of entanglement \cite{esb-1,esb-2,esb-3,esb-4}. In particular, it has been found that for a two-atom system with a nonvanishing separation immersed in a thermal bath, entanglement sudden birth only happens for atoms with an appropriate separation in a thermal bath at sufficiently small temperatures, while entanglement sudden death is a general feature \cite{esb-4}. However, when the interatomic separation is vanishing, entanglement can persist in the asymptotic equilibrium state depending on the initial state \cite{Benatti-job}. Here let us note that entanglement between atoms with nonvanishing separations may survive even in the asymptotic steady state when immersed in an environment out of thermal equilibrium \cite{bellomo-epl,bellomo-njp}.

A uniformly accelerated observer perceives the Minkowski vacuum as a thermal bath at a temperature proportional to its proper acceleration, which is the well-known Unruh effect \cite{Unruh}. Then a natural question is how the behaviors of entanglement between a pair of qubits are influenced by acceleration. Benatti and Floreanini have studied the entanglement generation for two uniformly accelerated atoms with a vanishing atomic separation and found that the asymptotic entanglement of such a two-atom system is exactly the same as that in a thermal bath at the Unruh temperature \cite{Benatti-pra}. Later, this work was generalized to the case of two accelerated atoms with a finite separation near a reflecting boundary, and it has been found that accelerated atoms may show distinct features from static ones in a thermal bath in terms of the entanglement creation in the neighborhood of the initial time \cite{yu-prd-07}. The studies mentioned above deal with entanglement either in the neighborhood of the beginning time or for the late equilibrium states, instead of the whole evolution process. Recently, the time evolution of entanglement for a two-qubit system has been investigated in Refs. \cite{Matsas-09,Doukas-09}, with the assumption that one of the qubits is accelerating while the other is inertial and isolated from the environment. In Ref. \cite{Doukas-10}, the authors study the entanglement dynamics of a two-qubit system accelerating at diametrically opposite points of a circular path initially in a Bell state,  assuming that the atoms are isolated from each other before adding the two together to solve for the total density operator.

In the present paper, we plan to study the entanglement dynamics of two mutually independent two-level atoms accelerating with the same acceleration perpendicular to the separation coupled with a bath of fluctuating massless scalar fields in the Minkowski vacuum. In particular, we investigate how entanglement decays for atoms initially prepared in a maximally entangled state and whether atoms in a separable initial state can get entangled during evolution. We also make a comparison between our results and those of static atoms immersed in a thermal bath at the Unruh temperature.


\section{The Master Equation}

We consider a two-atom system interacting with a bath of
fluctuating scalar fields in the Minkowski vacuum. The total Hamiltonian of such a system takes the form
\begin{equation}
 H=H_A+H_F+H_I\;.
\end{equation}
Here $H_A$ is the Hamiltonian of the two atoms,
\begin{equation}
H_A={\omega\over 2}\,\sigma_3^{(1)}+
    {\omega\over 2}\,\sigma_3^{(2)}\;,
\end{equation}
where $\sigma_i^{(1)}=\sigma_i\otimes{\sigma_0}$, $\sigma_i^{(2)}={\sigma_0}\otimes\sigma_i$, with $\sigma_i~(i=1,2,3)$ being the Pauli matrices and $\sigma_0$ being the $2\times2$ unit matrix. The two atoms share the same energy level spacing $\omega$. $H_F$ is the Hamiltonian of the scalar fields. In this paper, we aim to study the effects of vacuum fluctuations (modified by acceleration) on the dynamics of entanglement, so we assume that each of the two atoms interacts locally with a common bath of a fluctuating scalar field in the Minkowski vacuum and there are no direct interactions between the two atoms themselves. The interaction Hamiltonian $H_I$, which is supposed to be weak, is taken in analogy to the electric dipole interaction as \cite{Audretsch-94}
\begin{equation}
H_I=\mu\,[\sigma_{2}^{(1)} \Phi(t,{\bf x}_1)
    +\sigma_{2}^{(2)}\Phi(t, {\bf x}_2)\,]\;,
\end{equation}
where $\mu$ is the coupling constant.

We assume the atoms are uncorrelated with the environment at the beginning; that is, the initial state takes the form $\rho_{\text{tot}}(0)=\rho(0)\otimes\ket{0}\bra{0}$,
where $\ket{0}$ is the Minkowski vacuum state of the scalar fields, and $\rho(0)$ is the initial state of the two-atom system. In the weak-coupling limit, the reduced dynamics of the two-atom system takes the Kossakowski-Lindblad form \cite{Lindblad,Lindblad2,open}
\begin{equation}\label{master}
{\partial\rho(\tau)\/\partial\tau}
=-i\big[H_{\rm eff},\,\rho(\tau)\big]+{\cal L}[\rho(\tau)]\,,
\end{equation}
with
\begin{equation}
H_{\rm eff}
=H_A-\frac{i}{2}\sum_{\alpha,\beta=1}^2\sum_{i,j=1}^3
H_{ij}^{(\alpha\beta)}\,\sigma_i^{(\alpha)}\,\sigma_j^{(\beta)}\,,
\end{equation}
and
\begin{equation}
{\cal L}[\rho]
={1\over2} \sum_{\alpha,\beta=1}^2\sum_{i,j=1}^3
 C_{ij}^{(\alpha\beta)}
 \big[2\,\sigma_j^{(\beta)}\rho\,\sigma_i^{(\alpha)}
 -\sigma_i^{(\alpha)}\sigma_j^{(\beta)}\, \rho
 -\rho\,\sigma_i^{(\alpha)}\sigma_j^{(\beta)}\big]\,.
\end{equation}
Here $C_{ij}^{(\alpha\beta)}$ and
$H_{ij}^{(\alpha\beta)}$ are determined by the Fourier and Hilbert transforms, ${\cal G}^{(\alpha\beta)}(\lambda)$ and  ${\cal K}^{(\alpha\beta)}(\lambda)$, of the field correlation functions
\begin{equation}\label{green}
\mathrm{}G^{(\alpha\beta)}(\tau-\tau')
=\langle\Phi(\tau,\mathbf{x}_{\alpha})\Phi(\tau',\mathbf{x}_\beta)
 \rangle\;,
\end{equation}
 which are defined as
\begin{equation}\label{fourierG}
{\cal G}^{(\alpha\beta)}(\lambda)
=\int_{-\infty}^{\infty} d\Delta\tau\,
 e^{i{\lambda}\Delta\tau}\, G^{(\alpha\beta)}(\Delta\tau)\; ,
\end{equation}
\begin{equation}
{\cal K}^{(\alpha\beta)}(\lambda)
=\frac{P}{\pi i}\int_{-\infty}^{\infty} d\omega\
 \frac{{\cal G}^{(\alpha\beta)}(\omega)}{\omega-\lambda} \;,
\end{equation}
with $P$ denoting the principal value. Then $C_{ij}^{(\alpha\beta)}$ can be written explicitly as
\beq
C_{ij}^{(\alpha\beta)}
= A^{(\alpha\beta)}\delta_{ij}
 -iB^{(\alpha\beta)}\epsilon_{ijk}\,\delta_{3k}
 -A^{(\alpha\beta)}\delta_{3i}\,\delta_{3j}\;,
\eeq
where
\begin{equation}
\begin{aligned}
A^{(\alpha\beta)}
={\mu^2\/4}\,[\,{\cal G}^{(\alpha\beta)}(\omega)
 +{\cal G}^{(\alpha\beta)}(-\omega)]\;,\\
B^{(\alpha\beta)}
={\mu^2\/4}\,[\,{\cal G}^{(\alpha\beta)}(\omega)
 -{\cal G}^{(\alpha\beta)}(-\omega)]\;.
\end{aligned}
\end{equation}
Replacing ${\cal
G}^{(\alpha\beta)}$ with ${\cal
K}^{(\alpha\beta)}$ in the above equations, one obtains  $H^{(\alpha\beta)}_{ij}$.

\section{entanglement dynamics of the two-atom system}

In this section we investigate the dynamics of the two-atom system accelerating with the same acceleration perpendicular to the separation and compare it with that of static ones immersed in a thermal bath at the Unruh temperature.

The trajectories of the two uniformly accelerated atoms can be described as
\beq
\begin{aligned}\label{traj}
t_1(\tau)={1\/a}\sinh a\tau\;,~~~
x_1(\tau)={1\/a}\cosh a\tau\;,~~~
y_1(\tau)=0\;,~~~
z_1(\tau)=0\;,\\
t_2(\tau)={1\/a}\sinh a\tau\;,~~~
x_2(\tau)={1\/a}\cosh a\tau\;,~~~
y_2(\tau)=0\;,~~~
z_2(\tau)=L\;.
\end{aligned}
\eeq
The Wightman function of massless scalar fields in the Minkowski vacuum takes the form
\beq\label{wightman-m}
G^+(x,x')=-\frac{1}{4\pi^2}
           \frac{1}{(t-t'-i\epsilon)^2-(x-x')^2-(y-y')^2-(z-z')^2}\;.
\eeq
Allowing for the trajectories (\ref{traj}), the correlation functions can be written as
\beq
G^{(11)}(x,x')=G^{(22)}(x,x')
=-\frac{a^2}{16\pi^2}\frac{1}{\sinh^2(\frac{a(\tau-\tau')}{2}-i\epsilon)}\;,
\eeq
\beq
G^{(12)}(x,x')=G^{(21)}(x,x')
=-\frac{a^2}{16\pi^2}
  \frac{1}{\sinh^2(\frac{a(\tau-\tau')}{2}-i\epsilon)
  -{a^2L^2\/4}}\;.
\eeq
The Fourier transforms of the above correlation functions are
\beq
{\cal G}^{(11)}(\lambda)=G^{(22)}(\lambda)
=\frac{1}{2\pi}\frac{\lambda}{1-e^{-2\pi\lambda/a}}\;,
\eeq
\beq
{\cal G}^{(12)}(\lambda)=G^{(21)}(\lambda)
=\frac{1}{2\pi}\frac{\lambda}{1-e^{-2\pi\lambda/a}}
  f(\lambda,a,L)\;,
\eeq
where
\beq\label{f}
f(\lambda,a,L)=
  \frac{\sin \left(\frac{2\lambda}{a}
   \sinh^{-1}{a L\/2}\right)}
   {\lambda L\sqrt{1+{a^2 L^2/4}}}\;.
\eeq
Then the coefficients of the dissipator in the master equation can be obtained directly as
\begin{eqnarray}
&&C_{ij}^{(11)}=C_{ij}^{(22)}
=A_1\,\delta_{ij}-iB_1\epsilon_{ijk}\,\delta_{3k}
 -A_1\delta_{3i}\,\delta_{3j}\;,\\
&&C_{ij}^{(12)}=C_{ij}^{(21)}
=A_2\,\delta_{ij}-iB_2\epsilon_{ijk}\,\delta_{3k}
 -A_2\delta_{3i}\,\delta_{3j}\;,
\end{eqnarray}
where
\begin{eqnarray}\label{abc}
\begin{aligned}
&A_1={\Gamma_0\over4}\,\coth{\pi\omega\/a}\;,\\
&A_2={\Gamma_0\over4}\,f(\omega,a,L)\,
     \coth{\pi\omega\/a}\;,\\
&B_1={\Gamma_0\over4}\;,\\
&B_2={\Gamma_0\over4}\,f(\omega,a,L)\;,
\end{aligned}
\end{eqnarray}
with $\Gamma_0=\mu^2\omega/2\pi$ being the spontaneous emission rate.

Usually, the master equation is solved in the computational basis $\{\ket{1}=|00\rangle,\ket{2}=|10\rangle,\ket{3}=|01\rangle,\ket{4}=|11\rangle\}$. However, for certain cases, the coupled basis $\{|G\rangle=|00\rangle,|A\rangle={1\/\sqrt{2}}(|10\rangle-|01\rangle), |S\rangle={1\/\sqrt{2}}(|10\rangle+|01\rangle),|E\rangle=|11\rangle\}$ is more convenient. Then a set of equations describing the time evolution of the populations in the coupled basis, which are decoupled from other matrix elements, can be obtained as \cite{ent-states}
\bea
&\dot{\rho_G}=-4(A_1-B_1)\rho_G+2(A_1+B_1-A_2-B_2)\rho_A+2(A_1+B_1+A_2+B_2)\rho_S\;,\label{rho-g}\\
&\dot{\rho_A}=-4(A_1-A_2)\rho_A+2(A_1-B_1-A_2+B_2)\rho_G+2(A_1+B_1-A_2-B_2)\rho_E\;,\label{rho-a}\\
&\dot{\rho_S}=-4(A_1+A_2)\rho_S+2(A_1-B_1+A_2-B_2)\rho_G+2(A_1+B_1+A_2+B_2)\rho_E\;,\label{rho-s}\\
&\dot{\rho_E}=-4(A_1+B_1)\rho_E+2(A_1-B_1-A_2+B_2)\rho_A+2(A_1-B_1+A_2-B_2)\rho_S\;,\label{rho-e}
\eea
where $\rho_I=\langle I|\rho|I\rangle$, $I\in\{G,\;A,\;S,\;E\}$. Since $\rho_G+\rho_A+\rho_S+\rho_E=1$, only three of the above equations are independent.

We take concurrence \cite{concurrence} as a measurement of quantum entanglement, which is 1 for the maximally entangled states and 0 for separable states. For X states, namely, states with nonzero elements only along the diagonal and antidiagonal of the density matrix, the concurrence takes the form \cite{concurrence-x}
\beq
C[\rho(\tau)]=2\max\{0,K_1(\tau),K_2(\tau)\}\;,
\eeq
where
\beq
K_1(\tau)=|\rho_{23}(\tau)|-\sqrt{\rho_{11}(\tau)\rho_{44}(\tau)}\;,\quad
K_2(\tau)=|\rho_{14}(\tau)|-\sqrt{\rho_{22}(\tau)\rho_{33}(\tau)}\;,
\eeq
with $\rho_{ij}=\langle i|\rho|j\rangle$. In the following, we consider the entanglement dynamics for atoms initially prepared in states $|A\rangle$, $|S\rangle$, and $|E\rangle$. Since  the equations of motion of these populations are decoupled from other matrix elements, we have $\rho_{14}(\tau)=\rho_{AS}(\tau)=\rho_{GE}(\tau)=0$. Therefore, the concurrence can be expressed with the populations in the coupled basis as
\beq\label{concurrence}
C[\rho(\tau)]=\max\{0, K(\tau)\}\;,\quad
K(\tau)=|\rho_{S}(\tau)-\rho_{A}(\tau)|-2\sqrt{\rho_{G}(\tau)\rho_{E}(\tau)}\;.
\eeq

Before the investigation of the time evolution of entanglement, we first examine the behaviors of the asymptotic state by setting the rates of change of the populations in Eqs. (\ref{rho-g})-(\ref{rho-e}) equal to zero. For atoms with a nonvanishing separation (which ensures the coefficients of Eqs. (\ref{rho-g})-(\ref{rho-e}) different from 0), we find that
\beq
\rho_A(\infty)=\rho_S(\infty)=
\frac{e^{2\pi\omega/a}}{\left(e^{2\pi\omega/a}+1\right)^2}\;.\\
\eeq
Therefore, $K(\infty)$ (\ref{concurrence}) is negative, which implies that the accelerated atoms will get disentangled within a finite time. When the interatomic separation is vanishing, the asymptotic state depends on the initial state, so the atoms can be entangled \cite{Benatti-pra}. These conclusions are in agreement with  those in the thermal case \cite{esb-4,Benatti-job}.

Now we begin our  study of the entanglement dynamics for uniformly accelerated atoms with the same acceleration perpendicular to the separation. For atoms initially prepared in a maximally entangled state, we want to know how quantum entanglement between the two atoms decays, while for atoms initially in a separable state, we check whether they can get entangled during evolution.

A comparison between the modulating function $f(\omega,a,L)$ (\ref{f}) and that of the thermal case (see Eq. (37) in \cite {Benatti-job}) shows that the modulating function here depends on acceleration $a$, but the counterpart in the thermal case is temperature independent. Therefore, the entanglement dynamics for uniformly accelerated atoms would generally be different from the static ones immersed in a thermal bath at the Unruh temperature. We address these issues in detail in the following.

\subsection{ Atoms with maximally entangled initial states $\ket{A}$ and $\ket{S}$}

We begin our discussion with the cases when the two-atom system is initially prepared in the symmetric state $\ket{S}$ and the antisymmetric state $\ket{A}$, which are maximally entangled.

When the interatomic separation is very large ($L\to \infty$), the modulating function $f(\omega,a,L)$ tends to 0. For atoms initially in $\ket{A}$, the time evolution of the elements of the density matrix can be solved as
\bea
\begin{aligned}
&\rho_G(\tau)=
\frac{1}{(e^{2\pi\omega/a}+1)^2}
\left(-e^{2\pi\omega/a}e^{-8A_1\tau}-e^{2\pi\omega/a}(e^{2\pi\omega/a}-1)e^{-4A_1\tau}+e^{4\pi\omega/a}\right)\;,\\
&\rho_A(\tau)=
\frac{1}{(e^{2\pi\omega/a}+1)^2}
\left(e^{2\pi\omega/a}e^{-8A_1\tau}+(e^{4\pi\omega/a}+1)e^{-4A_1\tau}+e^{2\pi\omega/a}\right)\;,\\
&\rho_S(\tau)=
\frac{1}{(e^{2\pi\omega/a}+1)^2}
\left(e^{2\pi\omega/a}e^{-8A_1\tau}-2e^{2\pi\omega/a}e^{-4A_1\tau}+e^{2\pi\omega/a}\right)\;,\\
&\rho_E(\tau)=
\frac{1}{(e^{2\pi\omega/a}+1)^2}
\left(-e^{2\pi\omega/a}e^{-8A_1\tau}+(e^{2\pi\omega/a}-1)e^{-4A_1\tau}+1\right)\;.
\end{aligned}
\eea
Since the equations governing the time evolution of both $\rho_A$ and $\rho_S$ are the same in this limit, it is obvious that for atoms initially in $\ket{S}$, the solutions can be obtained by exchanging $\rho_A$ with $\rho_S$ in the above equations. Therefore, there is no difference in the dynamics of concurrence whether the initial state is $\ket{A}$ or $\ket{S}$ according to Eq. (\ref{concurrence}). Physically, the atoms can be regarded as being coupled to individual baths in the large separation limit. In this limit, the entanglement dynamics for the accelerated atoms and thermal ones can not be distinguished since the modulating functions for both cases take the same limiting value of 0. Our results are in agreement with those derived in Ref. \cite{Ali-09}, in which the entanglement dynamics for atoms immersed in individual reservoirs at finite temperatures are studied. Note the decay rate of concurrence at $\tau=0$ is $\Gamma_0\coth{\pi\omega\/2a}$, which is smaller than $\Gamma_0\coth{\pi\omega\/a}$, namely, the decay rate of $\rho_A$ or $\rho_S$.

When the interatomic separation is vanishingly small ($L\to0$), the atoms interact with the field modes in a collective and coherent way, which is usually referred to as the two-atom Dicke model \cite{dicke}. In this limit, the modulating function $f(\omega,a,L)$ (\ref{f}) tends to 1, and the coefficients $A_1=B_1$, $A_2=B_2$, so Eq. (\ref{rho-a}) reduces to $\dot{\rho}_A=0$; that is, the population of the antisymmetric state remains constant, which means that even the equilibrium state depends on the initial state. Here, for atoms initially in the antisymmetric state $\ket{A}$, the spontaneous transition is suppressed completely, and it is obvious that
\beq
\rho_A(\tau)=1\;,\quad\rho_G(\tau)=\rho_S(\tau)=\rho_E(\tau)=0\;,
\eeq
so the concurrence remains maximum during evolution. For atoms initially in the symmetric state $\ket{S}$, the transition rate of $\rho_S$ is enhanced as twice that of the large separation limit, and direct calculations show that
\bea
&&\rho_E(\tau)=
\frac{e^{\pi\omega/a}-1}{2(e^{2\pi\omega/a}-e^{\pi\omega/a}+1)}e^{-2\Gamma_1\tau}
-\frac{e^{\pi\omega/a}+1}{2(e^{2\pi\omega/a}+e^{\pi\omega/a}+1)}e^{-2\Gamma_2\tau}
+\frac{1}{e^{4\pi\omega/a}+e^{2\pi\omega/a}+1}\;,\nn\\
&&\rho_G(\tau)=
\frac{-e^{\pi\omega/a}(e^{\pi\omega/a}-1)}{2(e^{2\pi\omega/a}-e^{\pi\omega/a}+1)}e^{-2\Gamma_1\tau}
-\frac{e^{\pi\omega/a}(e^{\pi\omega/a}+1)}{2(e^{2\pi\omega/a}+e^{\pi\omega/a}+1)}e^{-2\Gamma_2\tau}
+\frac{e^{4\pi\omega/a}}{e^{4\pi\omega/a}+e^{2\pi\omega/a}+1}\;,\nn\\
&&\rho_S(\tau)=
\frac{(e^{2\pi\omega/a}-1)^2}{2(e^{2\pi\omega/a}-e^{\pi\omega/a}+1)}e^{-2\Gamma_1\tau}
+\frac{(e^{2\pi\omega/a}+1)^2}{2(e^{2\pi\omega/a}+e^{\pi\omega/a}+1)}e^{-2\Gamma_2\tau}
+\frac{e^{2\pi\omega/a}}{e^{4\pi\omega/a}+e^{2\pi\omega/a}+1}\;,\nn\\
&&\rho_A(\tau)=0\;,
\eea
where
\beq
\Gamma_1=\frac{e^{2\pi\omega/a}-e^{\pi\omega/a}+1}{e^{2\pi\omega/a}-1}\;\Gamma_0\;,\quad
\Gamma_2=\frac{e^{2\pi\omega/a}+e^{\pi\omega/a}+1}{e^{2\pi\omega/a}-1}\;\Gamma_0\;.
\eeq
So the atoms get disentangled within a finite time, and the decay rate of concurrence at $t=0$ is $2\Gamma_0\coth{\pi\omega\/2a}$, which is twice that of atoms with infinitely large separations. Since the modulating functions of both the accelerated atoms and the thermal ones take the same limiting value of 1 when the separation is vanishing, the entanglement dynamics for the two cases cannot be distinguished.

\begin{figure}[htbp]
\centering
\subfigure{\includegraphics[width=0.49\textwidth]{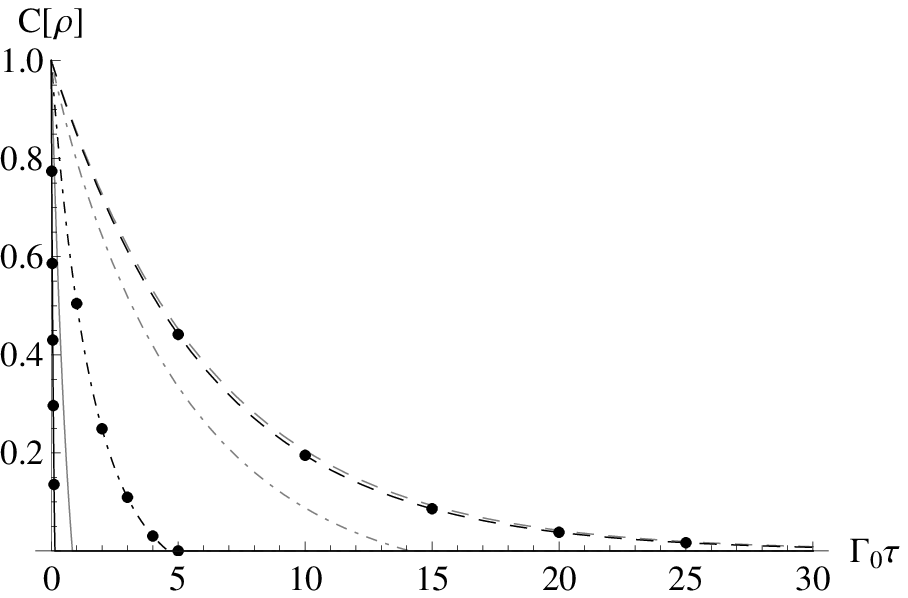}}
\subfigure{\includegraphics[width=0.49\textwidth]{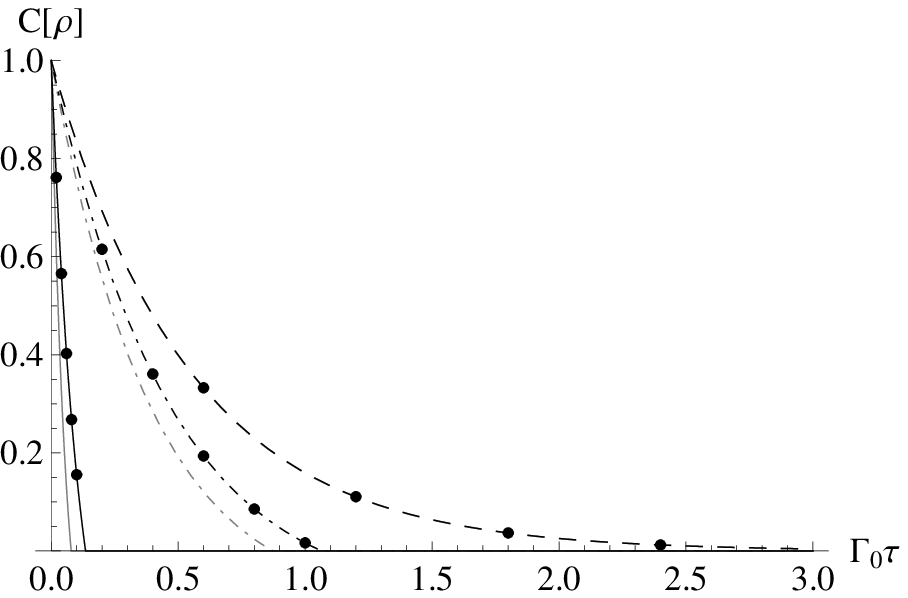}}
\caption{Comparison between the dynamics of concurrence for uniformly accelerated atoms (black lines marked with points) and static ones in a thermal bath at the Unruh temperature (gray lines) initially prepared in (left) $|A\rangle$ and (right) $|S\rangle$, with $\omega L=1$. The dashed, dot-dashed, and solid lines correspond to $a/\omega=2/10$, $a/\omega=2$, and $a/\omega=20$ respectively.}
\label{pic-l1}
\end{figure}
\begin{figure}[htbp]
\centering
\subfigure{\includegraphics[width=0.49\textwidth]{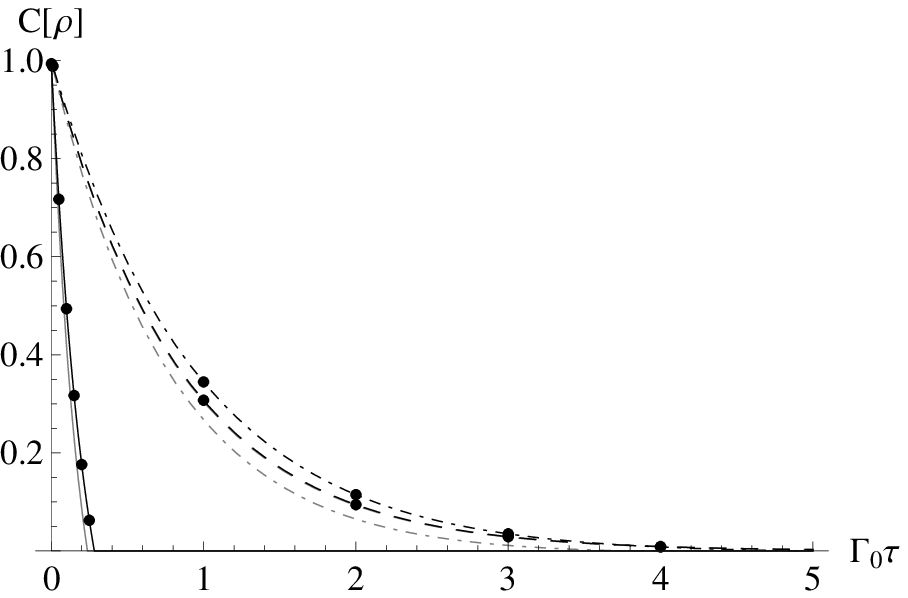}}
\subfigure{\includegraphics[width=0.49\textwidth]{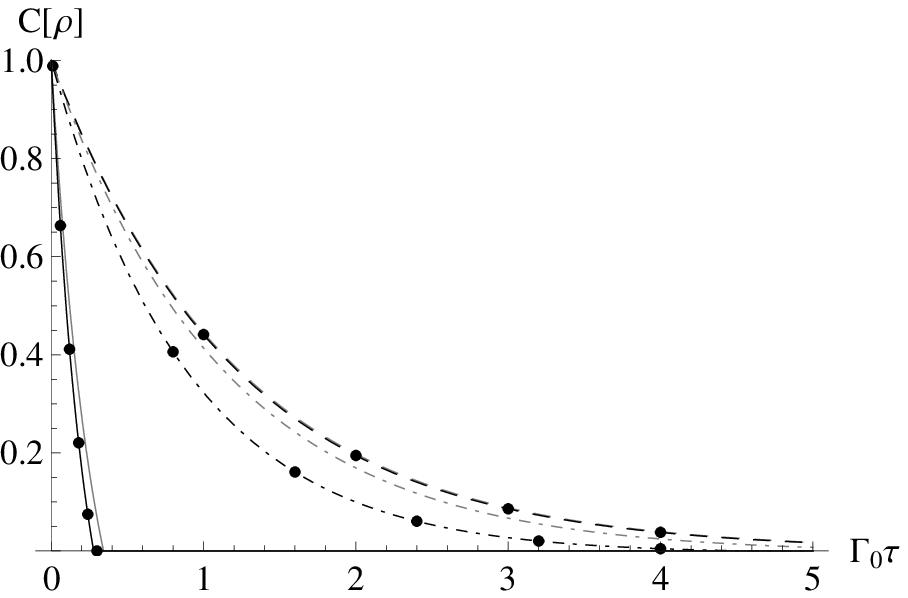}}
\caption{Comparison between the dynamics of concurrence for uniformly accelerated atoms (black lines marked with points) and static ones in a thermal bath at the Unruh temperature (gray lines) initially prepared in (left) $|A\rangle$ and (right) $|S\rangle$, with $\omega L=4$. The dashed, dot-dashed, and solid lines correspond to $a/\omega=1/10$, $a/\omega=1$, and $a/\omega=10$ respectively.}
\label{pic-l4}
\end{figure}

Now let us investigate the effects of acceleration on entanglement dynamics when the interatomic separation $L$ is comparable to the transition wavelength of the atoms $\sim\omega^{-1}$. In this regime, the solutions of Eqs. (\ref{rho-g})-(\ref{rho-e}) are rather complicated, so we solve these equations numerically. As discussed above, the decay rate of concurrence for atoms initially prepared in $\ket{A}$ and $\ket{S}$ is related to those of populations $\rho_A$ and $\rho_S$, which are proportional to $(1-f(\omega,a,L))$ and $(1+f(\omega,a,L))$ respectively. Since $f(\omega,a,L)$ is not a monotonic function of $a$ and $L$, the effect of acceleration $a$ on the decay rate of entanglement depends on the specific value of $L$. In the following, we study the effects of acceleration on entanglement dynamics for accelerated atoms initially prepared in $\ket{A}$ and $\ket{S}$ with two different interatomic separations, and compare the results with those of static ones in a thermal bath at the Unruh temperature in Figs. (\ref{pic-l1}) and (\ref{pic-l4}), respectively. In the case $\omega L=1$, the larger the acceleration is, the faster the concurrence decays. When $\omega L=4$, it is shown that, the decay rates may not increase with acceleration, which is distinct from the thermal case. For small accelerations, the entanglement dynamics of the uniformly accelerated atoms is essentially the same as that of the thermal case since the function $f(\omega,L)$ (\ref{f}) can be expanded with respect to the acceleration $a$ as
\beq\label{f-exp}
f(\omega,a,L)=
\frac{\sin\omega L}{\omega L}
-\frac{1}{24}\left(L^2{\cos\omega L}
+\frac{3L}{\omega}\,{\sin\omega L}\right)a^2
+O[a^4]\;,
\eeq
and the zeroth-order term takes exactly the same form as that in the thermal case. As the acceleration increases, the entanglement dynamics of the accelerated atoms becomes more distinguishable from that of the corresponding thermal case.

\subsection{ Atoms with a separable initial state $\ket{E}$}

Here, we investigate the entanglement dynamics for two-atom systems initially prepared in a separable state $|E\rangle$. From Eq. (\ref{concurrence}) we know that the factor $2\sqrt{\rho_G\rho_E}$ acts as a threshold, and thus, entanglement can be generated only if the difference of populations of the symmetric and antisymmetric states overweights this threshold. Therefore, when the atoms are initially in state $|E\rangle$, they may get entangled after a finite time of evolution via spontaneous emission \cite{esb-1}.

\begin{figure}[htbp]
\centering
\includegraphics[scale=0.8]{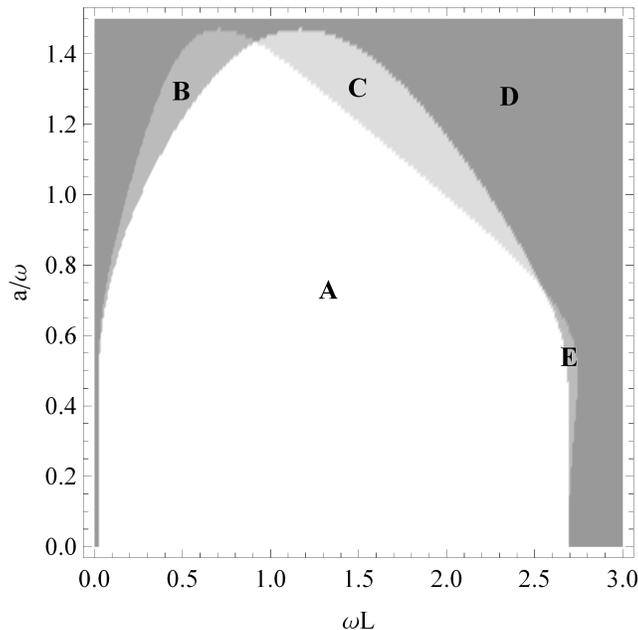}
\caption{ Entanglement profile for two-atom systems initially prepared in $\ket{E}$. Region A: both accelerated atoms and static ones in a thermal bath can get entangled. Region B: accelerated atoms can get entangled while static ones in a thermal bath cannot. Region C: accelerated atoms cannot get entangled while static ones in a thermal bath can. Region D: neither accelerated atoms nor static ones in a thermal bath can get entangled. Region E: accelerated atoms can get entangled while inertial ones in the Minkowski vacuum cannot.}
\label{phase-diag}
\end{figure}

Now let us investigate under what conditions entanglement sudden birth can happen for atoms initially prepared in $\ket{E}$. When the separation $L$ is vanishing, $A_1=A_2$, $B_1=B_2$, and $\rho_A$ remains zero during evolution. In this case, the threshold overweights the population $\rho_S$ all the time, and no quantum entanglement can be generated \cite{esb-1}. In addition, if the separation is very large, $A_2\approx 0$, $B_2\approx 0$, one derives from Eqs. (\ref{rho-a}) and (\ref{rho-s}) that ${d\/dt}(\rho_A-\rho_S)=-4A_1(\rho_A-\rho_S)$, thus $(\rho_A-\rho_S)$ remains 0 all the time for atoms initially prepared in $|E\rangle$, and no entanglement is created. In Fig. (\ref{phase-diag}) we study numerically the range of acceleration (temperature) within which entanglement can be generated for both accelerated atoms and static ones in a thermal bath initially prepared in $\ket{E}$ when $\omega L$ ranges from 0 to 3. Similar to the conclusion derived in Ref. \cite{esb-4} that entanglement sudden birth happens only when the temperature of the thermal bath is sufficiently small, here we find that for each interatomic separation, there exists an upper bound of acceleration larger than which entanglement cannot be generated. Another fact shown in this phase diagram is that the possible region of entanglement generation for accelerated atoms does not completely overlap with that for the static atoms in a thermal bath. That is, for certain circumstances, accelerated atoms can get entangled while the static ones in a thermal bath at the corresponding Unruh temperature cannot and vice versa. In particular, for certain interatomic separations, accelerated atoms with an appropriate acceleration can get entangled while the inertial ones in the Minkowski vacuum can not. Similar conclusions have been drawn in Ref. \cite{martin11}, in which it has been found that the degree of entanglement of some particular states shared by two observers increases as one of the observers accelerates, and in Ref. \cite{weak-gf}, in which the enhancement of vacuum entanglement by a weak gravitational field has been shown.

\begin{figure}[htbp]
\centering
\subfigure{\includegraphics[width=0.49\textwidth]{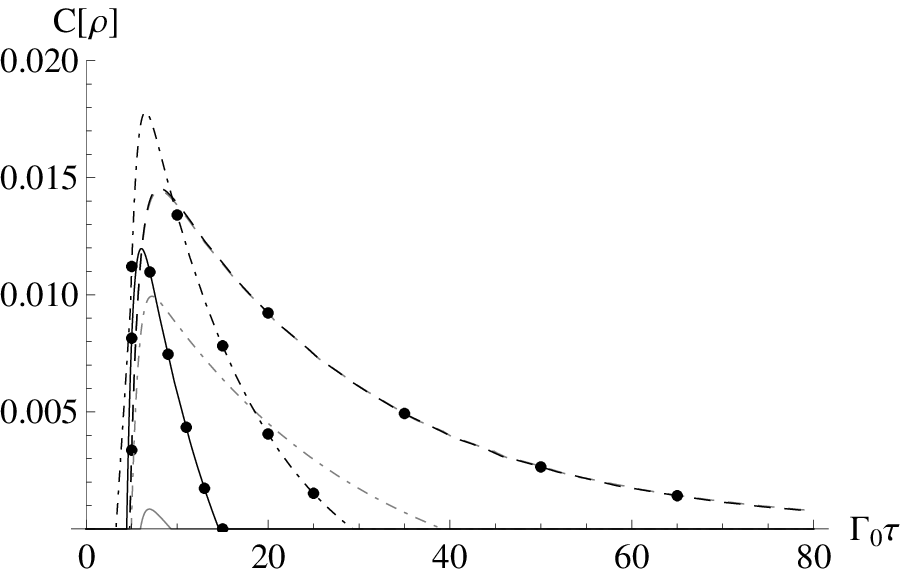}}
\subfigure{\includegraphics[width=0.49\textwidth]{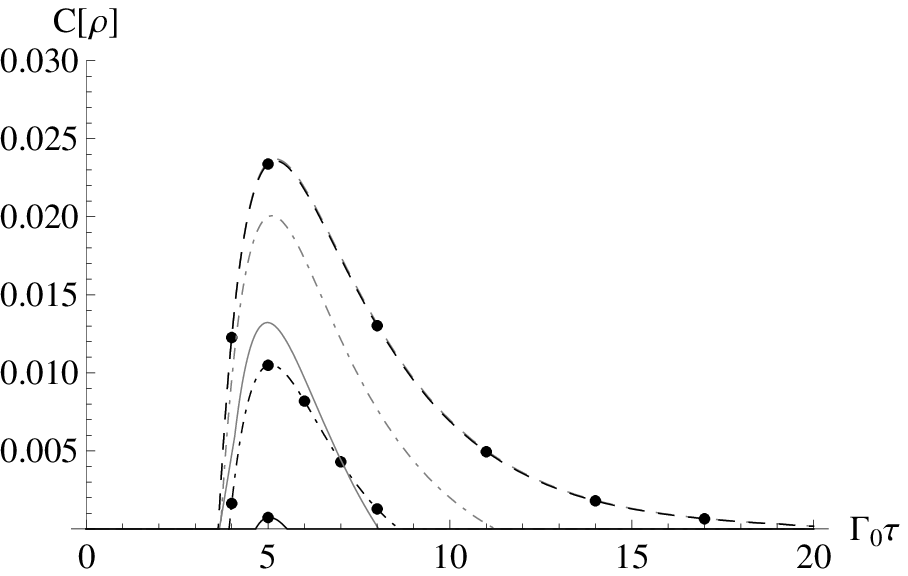}}
\caption{Comparison between the dynamics of concurrence for uniformly accelerated atoms (black lines marked with points) and static ones in a thermal bath at the Unruh temperature (gray lines) initially prepared in $|E\rangle$, with (left) $\omega L=1/2$ and (right) $\omega L=3/2$. The dashed, dot-dashed, and solid lines correspond to $a/\omega=1/10$, $a/\omega=1$, and $a/\omega=6/5$ respectively.}
\label{pic-E}
\end{figure}

In the following, we study the effects of acceleration on the evolution of concurrence for two-atom systems initially in $|E\rangle$ with two different interatomic separations in Fig. (\ref{pic-E}). It is shown that the lifetime of entanglement decreases as the acceleration grows. In the case $\omega L=3/2$, when the acceleration becomes larger than $6\omega/5$, entanglement generation does not happen for accelerated atoms, while the static ones in a thermal bath at the corresponding Unruh temperature can still be entangled. For atoms with separation $\omega L=1/2$, entanglement sudden birth for static atoms in a thermal bath stops first as the acceleration or the corresponding Unruh temperature increases.

\begin{figure}[htbp]
\centering
\subfigure{\includegraphics[width=0.49\textwidth]{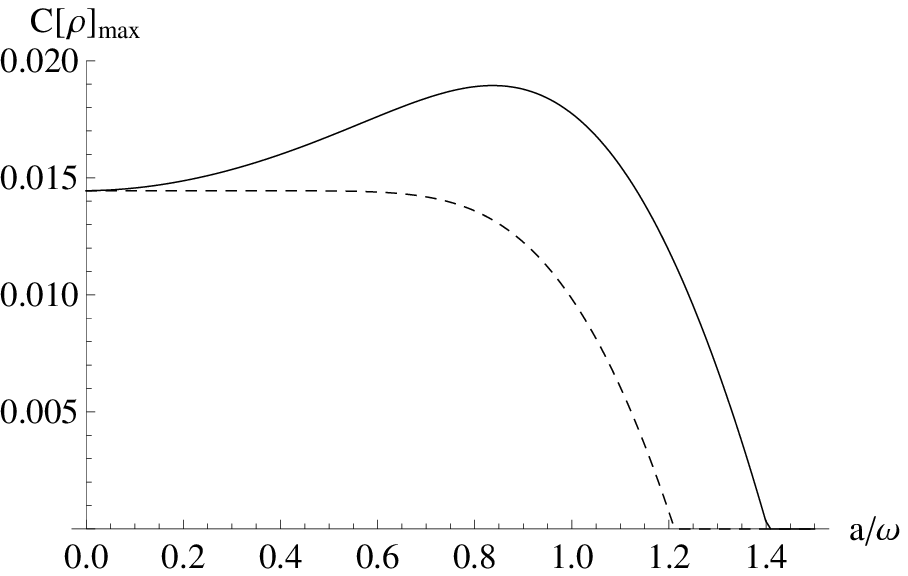}}
\subfigure{\includegraphics[width=0.49\textwidth]{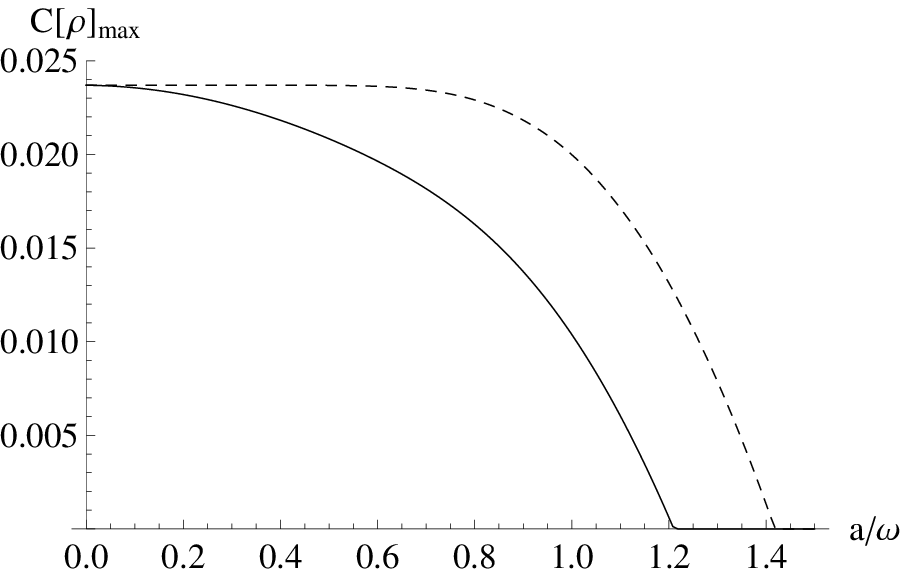}}
\caption{Comparison between the maximum of concurrence during evolution for uniformly accelerated atoms (solid lines) and static ones in a thermal bath at the Unruh temperature (dashed lines) initially prepared in $|E\rangle$ with (left) $\omega L=1/2$ and (right) $\omega L=3/2$.}
\label{pic-max-c}
\end{figure}

Another point we want to address is the maximum of entanglement generated during evolution. Intuitively, one may expect it would decrease as the acceleration increases as a result of the Unruh effect. However, we find that this is not always the case. It is shown in Fig. (\ref{pic-max-c}) that, for the thermal case, the maximum of concurrence always decreases as the temperature increases. When the temperature is small, the maximum of concurrence varies extremely slow with acceleration and is almost a constant. However, for the accelerated ones, this maximum may not decrease with acceleration for certain separations. In particular, it may even exceed that of statics ones in vacuum. In the following, we give a brief approximate analysis of how this happens when the acceleration is small. In the limit of small acceleration or temperature, the spontaneous excitations can be neglected, and the factor $\coth{\pi\omega\/a}$ can be approximated as 1. In fact, $\coth{\pi\omega\/a}-1$ is an infinitesimal of higher order than $a^N$, with $N$ being any finite natural number. This leads to $A_1-B_1=0$ and $A_2-B_2=0$, and then the time evolution of the populations (\ref{rho-g})-(\ref{rho-e}) can be solved analytically as
\bea
\begin{aligned}
&\rho_G(\tau)=
1-\frac{1+f}{1-f}e^{-(1+f)\Gamma_0\tau}
-\frac{1-f}{1+f}e^{-(1-f)\Gamma_0\tau}
+\frac{1+3f^2}{1-f^2}e^{-2\Gamma_0\tau}\;,\\
&\rho_A(\tau)=
\frac{1-f}{1+f}
e^{-2\Gamma_0\tau}\left(e^{(1+f)\Gamma_0\tau}-1\right)\;,\\
&\rho_S(\tau)=
\frac{1+f}{1-f}
e^{-2\Gamma_0\tau}\left(e^{(1-f)\Gamma_0\tau}-1\right)\;,\\
&\rho_E(\tau)=
e^{-2\Gamma_0\tau}\;,
\end{aligned}
\eea
with $f$ being the modulating function. Then, for the thermal case, $f(\omega,L)$ is temperature independent, so the concurrence is also independent of temperature in this approximation, which explains why the maximum of concurrence is almost a constant for small accelerations. As for the accelerated case, the maximum of concurrence can either increase or decrease with acceleration depending on the specific value of $L$, as $f(\omega,a,L)$ is acceleration dependent.


\section{Conclusion}

In conclusion, we have studied the dynamics of two uniformly accelerated two-level atoms in the Minkowski vacuum in the framework of open quantum systems. We take concurrence to be a measurement of quantum entanglement and investigate how it evolves. For atoms in a maximally entangled state, entanglement sudden death is a general feature for accelerated atoms with a nonvanishing separation. In contrast to the thermal case, the decay rate of entanglement may not necessarily increase with acceleration. When both of the two accelerated atoms are initially in the excited state, the conditions for entanglement generation are investigated numerically and are found not to be completely the same as those for static ones in a thermal bath. Remarkably, for certain interatomic separations, accelerated atoms can get entangled while the inertial ones in the Minkowski vacuum can not, and the maximum of concurrence generated during evolution for accelerated atoms may increase with acceleration.

\begin{acknowledgments}
This work was supported in part by the NSFC under Grants No. 11375092, No. 11435006, and No. 11447022; the SRFDP under Grant No. 20124306110001; the Research Program of Ningbo University
under Grant No. XYL15020; and the K. C. Wong Magna Fund in Ningbo University.
\end{acknowledgments}

\end{document}